\documentclass[10pt,twocolumn,english,twocolumn,prl]{revtex4-1}
\usepackage[T1]{fontenc}
\usepackage[latin9]{inputenc}
\usepackage{xcolor}
\usepackage{amsmath}
\usepackage{graphicx}
\PassOptionsToPackage{normalem}{ulem}
\usepackage{ulem}

\makeatletter

\providecommand{\tabularnewline}{\\}
\providecolor{lyxadded}{rgb}{0,0,1}
\providecolor{lyxdeleted}{rgb}{1,0,0}

\usepackage{lmodern}
\usepackage[T1]{fontenc}

\makeatother

\usepackage{babel}
\begin{document}

\title{Coupling ideality of integrated planar high-Q microresonators}

\author{Martin H. P. Pfeiffer\textsuperscript{\dag{}}, Junqiu Liu\textsuperscript{\dag{}},
Michael Geiselmann\textsuperscript{}, Tobias J. Kippenberg\textsuperscript{}}
\email{tobias.kippenberg@epfl.ch}

\affiliation{École Polytechnique Fédérale de Lausanne (EPFL), CH-1015 Lausanne,
Switzerland}

\collaboration{\dag{} These authors contributed equally to this work.}

\begin{abstract}
Chipscale microresonators with integrated planar optical waveguides
are useful building blocks for linear, nonlinear and quantum optical
devices. Loss reduction through improving fabrication processes has
resulted in several integrated microresonator platforms attaining
quality (Q) factors of several millions. However only few studies
have investigated design-dependent losses, especially with regard
to the resonator coupling section. Here we investigate
design-dependent parasitic losses, described by the coupling ideality,
of the commonly employed microresonator design consisting of a microring
resonator waveguide side-coupled to a straight bus waveguide. By systematic
characterization of multi-mode high-Q silicon nitride microresonator
devices, we show that this design can suffer from low coupling ideality.
By performing full 3D simulations to numerically investigate the resonator
to bus waveguide coupling, we identify the coupling to higher-order
bus waveguide modes as the dominant origin of parasitic losses which
lead to the low coupling ideality. Using suitably designed bus waveguides,
parasitic losses are mitigated, and a nearly unity ideality and strong
overcoupling (i.e. a ratio of external coupling to internal resonator
loss rate $>9$) are demonstrated. Moreover we find that different
resonator modes can exchange power through the coupler, which therefore
constitutes a mechanism that induces modal coupling, a phenomenon
known to distort resonator dispersion properties. Our results demonstrate
the potential for significant performance improvements of integrated
planar microresonators, achievable by optimized coupler designs.
\end{abstract}

\maketitle

\section{Introduction}

Microresonator devices are ubiquitously used in integrated photonic
circuits and enable applications that range from passive elements
such as filters \citep{Little1997} and sensors \citep{DeVos2007},
over active components such as modulators \citep{Reed2005}, to nonlinear
applications \citep{Leuthold2010,Moss2013} such as wavelength conversion
\citep{Li2015} and Kerr frequency comb generation \citep{Kippenberg2011}.
While most microresonator devices in silicon photonics are formed
by single-mode waveguides \citep{Jalali2006,Nagarajan2005}, many
recent photonic integrated circuits rely on multi-mode waveguides
due to their lower losses \citep{Cherchi2013,Bauters2011}, higher
data capacity \citep{Luo2014}, improved device integration \citep{Wade2014}
and tailored dispersion properties e.g. to attain anomalous group
velocity dispersion required for parametric frequency conversion \citep{Turner2006,Riemensberger2012}.
Early research on ultra high-Q microresonators in other platforms
led to the development of several adjustable evanescent coupling techniques
based on prisms and tapered optical fibers \citep{Braginsky1989,Knight1997,Gorodetsky1999,Cai2000,Spillane2003}.
To quantitatively describe the performance of these couplers , the
``coupling ideality'' was defined for tapered fiber coupling to
microspheres, as the ratio of the power coupled from the resonator
to the fundamental fiber mode divided by the total power coupled to
all guided and non-guided fiber modes \citep{Spillane2003}. In this
case parasitic losses degrading coupling ideality can be present if
the tapered fiber is multi-mode.

\begin{figure}[!t]
\includegraphics[width=1\columnwidth]{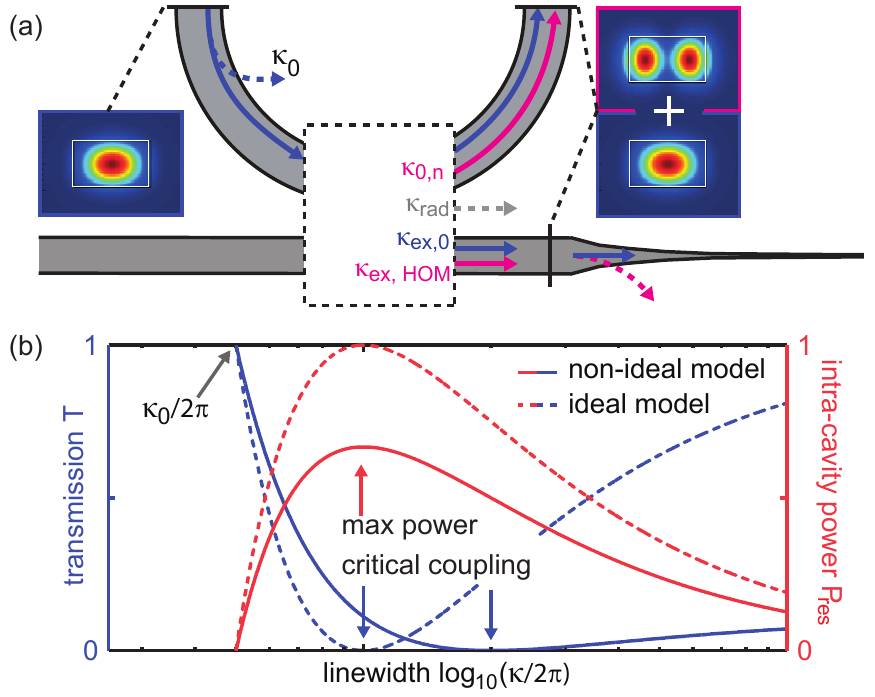}
\caption{(a) Schematic representation of the coupling rates in an integrated
microresonator with multi-mode waveguides. The parasitic coupling
processes of a fundamental resonator mode are illustrated. $\kappa_{0}$ represents the resonator's internal
loss rate and $\kappa_{\mathrm{ex},0}$ represents the coupling rate
to the fundamental bus waveguide mode. $\kappa_{\mathrm{ex,HOM}}$
represents the coupling rate to the higher-order bus waveguide modes,
which are later filtered out by the inverse taper mode converter.
$\kappa_{0,n}$ represents the coupling rate to other resonator modes,
while $\kappa_{\mathrm{rad}}$ represents the coupling rate to free
space modes. (b) Plot of the transmission $T$ (blue) and the intra-cavity
power $P_{\mathrm{res}}$ (red) as function of the total linewidth
$\kappa/2\pi$ for the ideal ($I=1$, dashed lines) and non-ideal
($I=0.67$, solid lines) case. }
\label{fig1}
\vspace{-0.2cm}
\end{figure}

In the context of integrated planar microresonator devices, design
rules \citep{Soltani2007,Hosseini2010} and optimized coupler geometries
\citep{Ghulinyan2013,Pencer2014,Wade2014} have been reported. However,
comparatively little attention has been paid to the coupler performance,
especially with regard to the multi-mode nature of waveguides. Only
few reports of coupler-induced excess losses \citep{Fengnian2006,Li2016}
have been published and most integrated microresonator devices, single-
or multi-mode, rely on the coupler design consisting of a simple side-coupled
straight bus waveguide with a cross section identical to the resonator
waveguide. 

Here we present a comprehensive investigation of integrated planar
high-Q silicon nitride ($\mathrm{Si_{3}N_{4}}$) microresonator devices
with several different coupler designs. Experimental resonance characterization
with sufficiently large statistics and full 3D numerical simulations
allow us to unambiguously reveal the detrimental effect of non-ideal
coupler designs, even in the presence of statistical fluctuations
of resonator properties due to fabrication variations. The commonly
employed coupler design using a bus waveguide of the same cross
section as the resonator is found to exhibit parasitic losses due
to the modal coupling to higher-order bus waveguide modes, which can
severely limit the device performance. In contrast, for the design
of a multi-mode resonator coupled to a single-mode bus waveguide,
we observe nearly ideal coupler performance. Finally, our simulations
show that coupling between different resonator modes can originate
from the coupler. This provides a novel insight into the origin of
modal coupling in microresonators observed in previous work \citep{Herr2014,Braschaad4811},
which leads to distortion of resonator dispersion properties.

\section{Analytical description of a multi-mode coupling section}

Typically the evanescent coupling of light to a microresonator is
described using coupled-mode theory as a power transfer to a resonator
mode at the rate $\kappa_{\mathrm{ex,0}}$ \citep{Gorodetsky1999,Cai2000,haus1984waves}.
Treating the resonator in a lumped model \citep{Rowland1993,Yariv2000},
the coupling rate $\kappa_{\mathrm{ex},0}$ is typically estimated
using the model of coupling between two co-propagating modes in adjacent
waveguides \citep{Yariv1973}. In contrast to the power coupling ratios
of conventional directional couplers, the high-Q microresonator's
low internal loss rate $\kappa_{0}$ requires only minute power transfer
to achieve critical coupling (i.e. $\kappa_{\mathrm{ex,0}}=\kappa_{0}$
) for which the intra-cavity power build-up is maximal. Thus the coupled
modes in both the resonator and the bus waveguides can be essentially
treated as independent, and $\kappa_{\mathrm{ex,0}}$ depends on the
mutual modal overlap and propagation constant mismatch
(i.e. phase mismatch) \citep{Little1997,Gorodetsky1999,Yariv1973}.
This model is widely applied as it provides a qualitative insight
for most cases where coupling between only two modes is considered,
neglecting the coupling to other modes.

In practice for high-Q microresonators, a commonly employed coupler
design consists of a side-coupled, straight bus waveguide identical
in cross section to the resonator waveguide. The cross section is
chosen in order to match the propagation constants of e.g. the fundamental
resonator and bus waveguide modes. However in the case of multi-mode
waveguides, as found for tapered fiber coupling to microspheres
\citep{Spillane2003}, coupling between different modes has to be
considered as depicted in Fig. \ref{fig1}(a). Moreover, the coupler
can scatter light into free space modes and recently was also identified
to couple the counter-propagating, clockwise (CW) and counter-clockwise
(CCW) waveguide modes \citep{Li2016}, which is not considered in
the present work. As a result the corresponding equation of motion
for the resonator modal amplitude $a_{0}$ of frequency $\omega_{0}$
in the rotating frame of the driving laser $\omega_{L}$ has to be
extended to: 

\begin{multline}
\frac{d}{dt}a_{0}=i\varDelta_{0}a_{0}-\left(\frac{\kappa_{0}+\kappa_{\mathrm{ex},0}+\kappa_{\mathrm{p}}}{2}\right)a_{0}\\
+\sqrt{\kappa_{\mathrm{ex},0}}s_{\mathrm{in}}+\frac{i}{2}\sum_{n\neq0}\kappa_{0,n}a_{n}e^{i\Delta_{n}t}
\end{multline}

Here $\varDelta_{0}=\omega_{L}-\omega_{0}$ and $\Delta_{n}=\omega_{L}-\omega_{n}$
are the frequency detunings between the driving laser with amplitude
$s_{\mathrm{in}}$ and the resonator modes $a_{0}$ and $a_{n}$.
The intra-cavity field decays due to the internal loss rate $\kappa_{0}$
and the external coupling rate $\kappa_{\mathrm{ex},0}$ to the fundamental
bus waveguide mode. The radiations into free space modes with the rate
$\kappa_{\mathrm{rad}}$ and to higher-order bus waveguide modes with
the rate $\kappa_{\mathrm{ex,HOM}}=\sum_{q\neq0}\kappa_{\mathrm{ex,\mathit{q}}}$ form the parasitic
coupling rate $\kappa_{\mathrm{p}}=\kappa_{\mathrm{rad}}+\kappa_{\mathrm{ex,HOM}}$,
which accelerates the intra-cavity field decay.

In addition, the modal coupling term $\frac{i}{2}\sum_{n\neq0}\kappa_{0,n}a_{n}e^{i\Delta_{n}t}$
is introduced to account for the fact that the resonator mode $a_{0}$
can couple to other modes with the rate $\kappa_{0,n}$. Such modal
coupling is usually considered to arise from surface roughness, but
is later found to originate also from the coupler. This term is only
relevant if the coupled modes are simultaneously resonant. Such modal
coupling causes deviations of the resonance frequencies, so called
``avoided modal crossings'', that locally distort resonator dispersion.
As the modal coupling term contributes to the parasitic coupling rate $\kappa_{\mathrm{p}}$
only at such modal crossing points, it is not included in $\kappa_{\mathrm{p}}$.
The coupling ideality $I$ of the resonator mode $a_{0}$, describing
the relative strength of parasitic coupling rates, is defined according
to Ref. \citep{Spillane2003} as:

\begin{equation}
I=\frac{\kappa_{\mathrm{ex},0}}{\kappa_{\mathrm{ex},0}+\kappa_{\mathrm{p}}}\label{eq:1}
\end{equation}

In the following the effects of coupling ideality on device performance
are considered. While the scattering of light into free space modes
directly represents a power loss, the power coupled to higher-order modes
of the bus waveguide is not necessarily lost. However in most cases
the higher-order bus waveguide modes are filtered out e.g. by inverse
taper mode converters \citep{Almeida2003}. Thus the measured transmitted power
at the device facets only consists of the power of the bus waveguide's
fundamental mode, and the input-output relation $s_{\mathrm{out}}=s_{\mathrm{in}}-\sqrt{\kappa_{\mathrm{ex},0}}a_{0}$
holds with $\kappa_{\mathrm{ex,HOM}}$ representing a parasitic loss
which enlarges the resonance linewidth. On resonance $\varDelta_{0}=0$,
the device power transmission $T$ and intra-cavity power $P_{\mathrm{res}}$
as function of the coupling ideality $I$ and coupling parameter $K\text{=}\kappa_{\mathrm{ex},0}/\kappa_{0}$
are expressed as:

\begin{equation}
T=|1-\frac{2}{K^{-1}+I^{-1}}|^{2}\label{eq:2}
\end{equation}

\begin{equation}
P_{\mathrm{res}}=\frac{D_{1}}{2\pi}\cdot\frac{4}{\kappa_{\mathrm{ex},0}(K^{-1}+I^{-1})^{2}}P_{\mathrm{in}}
\end{equation}

Here $D_{1}/2\pi$ is the resonator free spectral range (FSR). Assuming
an input power $\mathrm{\mathit{P}_{in}=}|s_{\mathrm{i\mathrm{n}}}|^{2}=1$
and a constant $D_{1}$, Fig. \ref{fig1}(b) plots both the power transmission
$T$ and intra-cavity power $\mathrm{\mathit{P}_{res}}$ as function
of the total linewidth $\kappa/2\pi=(\kappa_{0}+\kappa_{\mathrm{ex,0}}/I)/2\pi$
for the ideal ($I=1$) and non-ideal ($I=0.67$) case, with a constant
$\kappa_{0}$ and varying $\kappa_{\mathrm{ex,0}}$. The effects of
the non-ideal coupling become apparent: in the case of the ideal coupling
(dashed lines), the point of the full power extinction (i.e. $\mathrm{\mathit{T}}=0$,
the critical coupling point) coincides with the point of the
maximum intra-cavity power. This is different for the non-ideal case
(solid lines), in which the parasitic losses increase linearly with
the coupling rate $\kappa_{\mathrm{p}}=0.67\kappa_{\mathrm{ex,0}}$.
More importantly the maximum value of the intra-cavity power is reduced
compared to the ideal case. Due to the parasitic losses, critical
coupling and overcoupling are only achieved at larger total linewidth, or can not be achieved at all if $\kappa_{\mathrm{p}}>\kappa_{\mathrm{ex},0}$. It is therefore evident that in applications exploiting the resonator's
power enhancement e.g. nonlinear photonics, device performance
will improve with higher coupling ideality. Likewise, the analysis
shows that linewidth measurements carried out near the critical coupling
point include possible parasitic losses, preventing faithful
measurements of the intrinsic quality factor.

\begin{figure*}[t!]
\includegraphics[]{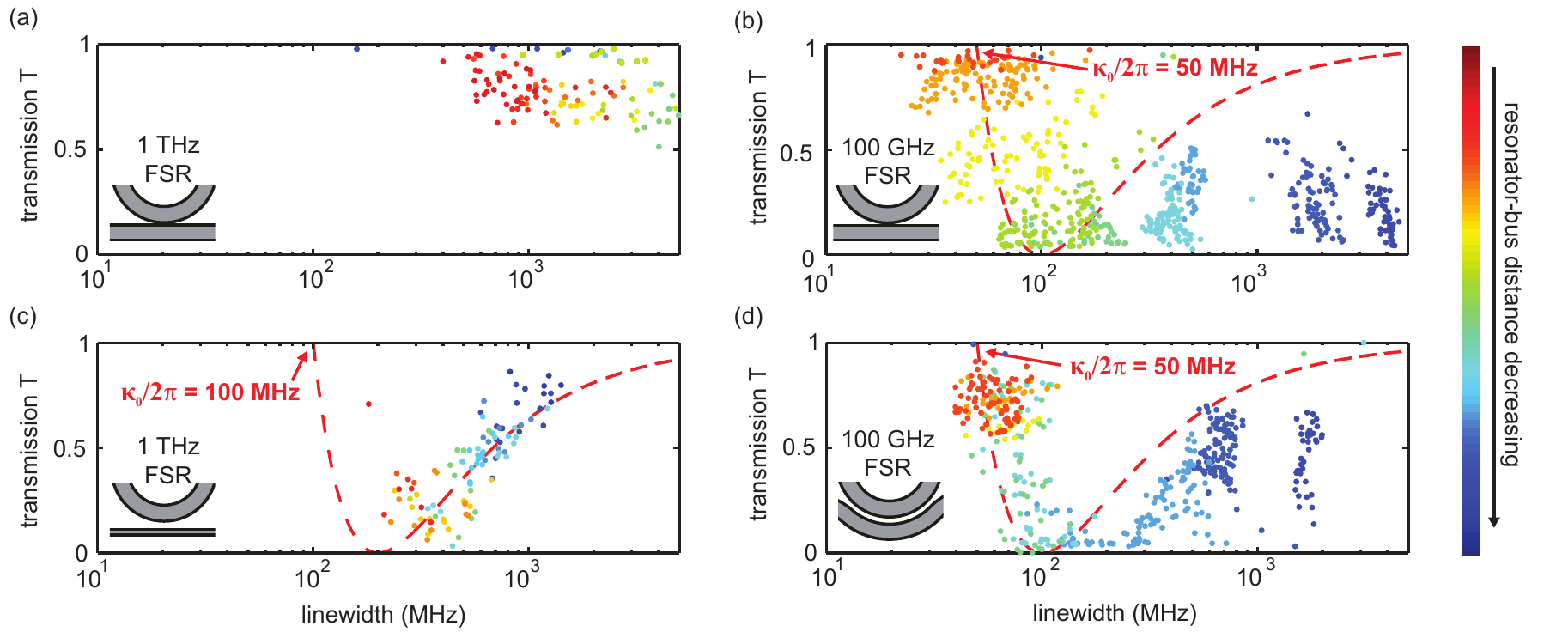}
\caption{Characterization of coupling ideality for the fundamental $\mathrm{TM_{R,00}}$ mode family of 1 THz FSR (Panels a, c) and 100 GHz FSR (Panels b,
d) microresonator devices. Dots of the same color correspond to resonances
of the same resonator with the color bar indicating their mutual resonator-bus
distance trend. The red dashed lines indicate the trend expected for
ideal coupling to resonators with the internal loss rates of $\kappa_{0}/2\pi=100$ MHz
(c) and $\kappa_{0}/2\pi=50$ MHz (b, d). For (a), (b) and
(d) the bus waveguide has the same cross section as the resonator
waveguide. Panels (b) and (d) show improved ideality and achieved
overcoupling through the use of a single-mode bus waveguide (b) and
a pulley-style coupler (d).}
\label{fig2}
\end{figure*}

\section{Experimental study of coupling ideality}

We experimentally study the coupling ideality for integrated $\mathrm{Si_{3}N_{4}}$
microresonators, a widely employed platform for on-chip nonlinear
photonics such as Kerr frequency comb generation \citep{Kippenberg2011}
and soliton formation \citep{Herr2013a}. For microresonator platforms
with adjustable couplers e.g. tapered fibers and prism couplers, changing
the evanescent coupling rates allows to measure the transmission-linewidth
dependence of a single resonance \citep{Spillane2003,Gorodetsky1999}
and to retrieve the coupling ideality via Eq. \ref{eq:2}. In contrast,
here we study photonic chips with several microresonator devices that
consist of resonator and bus waveguides, as well as inverse taper
mode converters \citep{Almeida2003} placed at the chip facets. The
microresonator devices on each chip are identical but have varying
resonator-bus distances providing different
coupling rates. In this case, coupling ideality is evaluated by analyzing
the transmission-linewidth dependence of many resonances acquired
for each microresonator device. By using a statistically large number
of devices we overcome the variations in quality factor Q inherent
to the fabrication process itself. 

The waveguide core is made from silicon nitride ($\mathrm{Si_{3}N_{4}}$)
and fully cladded with silicon dioxide ($\mathrm{SiO_{2}}$). All
measured chips were fabricated on the same wafer using a photonic
Damascene process \citep{Pfeiffer2016}. In contrast to typical subtractive
processes, this process allows for void-free, high-aspect-ratio coupler
gap fabrication, eliminating excess losses due to the presence of
voids. By using lensed fibers, light is coupled efficiently (loss$<3\mathrm{dB}$
per facet) into a single fundamental mode of the bus waveguide. Calibrated
power transmission traces are acquired for all devices on the chip from
1500 nm to 1630 nm with a similar method as described in Ref. \citep{DelHaye2009}.
A polarization controller is used to select and maintain a stable
input polarization over the full measurement bandwidth. Resonances
in each recorded device transmission trace are automatically identified
and fitted using a model of a splitted Lorentzian lineshape \citep{Gorodetsky2000}.
The resonances are grouped into different mode families by measuring
their mutual FSRs and comparing them to finite-element simulations
of the device geometry.

Fig. \ref{fig2} compares the measured transmission-linewidth dependence
of the resonator's transverse magnetic fundamental mode families ($\mathrm{TM_{R,00}}$) for two 1 THz FSR (Panels a, c) and two 100 GHz FSR (Panels b, d)
microresonator device chips. The cross section of the resonator waveguide
is $0.87$ $\mathrm{\mu m}$ height, and $2$ $\mathrm{\mu m}$ (100
GHz FSR) and $1.5$ $\mathrm{\mu m}$ (1 THz FSR) width respectively.
Each point represents a measured resonance, and points with the
same color are from the same microresonator device. Different colors
denote different resonator-bus distances.
The red dashed line traces out the transmission-linewidth dependence
for the ideal coupling of unity ideality with a fixed internal loss
$\kappa_{0}$.

Fig. \ref{fig2}(a) shows an example of low coupling ideality: a small
radius ($r\approx23$ $\mathrm{\mu m}$), 1 THz FSR resonator coupled
to a multi-mode bus waveguide of the same cross section. The measured
resonances of the fundamental $\mathrm{TM_{R,00}}$ mode family have
GHz linewidth and low extinction (i.e. high transmission), and their
measured transmission-linewidth dependence does not follow a clear
trend. Due to the identical cross sections of the resonator and the
bus waveguides, this coupler design could be naively assumed to provide
good propagation constant match between the resonator and bus waveguide
TM fundamental modes, i.e. $\mathrm{TM_{R,00}}$ and $\mathrm{TM_{B,00}}$.
However due to the small ring radius,
the propagation constants of the $\mathrm{TM_{R,00}}$ and $\mathrm{TM_{B,00}}$
modes are strongly mismatched, despite the identical waveguide cross
sections. 

As shown in Fig. \ref{fig2}(b), also a 100 GHz FSR resonator, with
a ten times larger radius ($r\approx230\mathrm{\mu m}$), can have
limited coupling ideality when interfaced with a straight bus waveguide
of the same cross section. Although featuring resonance linewidths
below $\kappa_{0}/2\pi=30$ MHz and an average linewidth of $\kappa_{0}/2\pi\approx50$
MHz, the microresonator can not be efficiently overcoupled, indicating
the presence of parasitic losses.

Fig. \ref{fig2}(c) and (d) present two possible coupler designs that
improve coupling ideality. First, as shown in Fig. \ref{fig2}(c)
almost unity ideality and strong overcoupling are achieved for a 1
THz FSR microresonator coupled to a single-mode bus waveguide. The
bus waveguide has a cross section of $0.6$ $\mathrm{\mu m}$ height
and $0.4$ $\mathrm{\mu m}$ width due to the aspect-ratio-dependent
etch rate during the preform etch \citep{Pfeiffer2016}. It can thus
be concluded that the main source of parasitic losses leading to the
low ideality in Fig. \ref{fig2}(a) originates from the coupling to
higher-order bus waveguide modes. Therefore using a single-mode bus
waveguide can essentially avoid this kind of parasitic losses and
significantly improve coupling ideality to near unity. Also
strong overcoupling can be achieved with an external coupling rate
$\kappa_{\mathrm{ex,0}}$ almost a magnitude larger than the internal
losses (coupling parameter $K\text{=}\kappa_{\mathrm{ex},0}/\kappa_{0}=\kappa/\kappa_{0}-1>9$).

However in most cases when using a single-mode bus waveguide, though
coupling ideality is improved, the propagation constants of the bus and
resonator fundamental modes (e.g. $\mathrm{TM_{B,00}}$ and $\mathrm{TM_{R,00}}$)
are strongly mismatched which limits the maximum value of the coupling
rate $\kappa_{\mathrm{ex,0}}$. Thus a narrow gap is needed to achieve
sufficient modal overlap and a large enough coupling rate $\kappa_{\mathrm{ex,0}}$
to achieve overcoupling. For the 1 THz FSR resonator, a coupling rate
$\kappa_{\mathrm{ex,0}}$ sufficient for overcoupling is achieved
due to its small mode volume and low internal loss per round-trip
($\propto\kappa_{0}/D_{1}$). However for smaller FSR resonators with
larger mode volumes e.g. 100 GHz FSR, overcoupling might not be achieved
in the case of strong propagation constant mismatch, as fabrication
processes pose limitations on the narrowest resonator-bus distance.
One alternative solution for smaller FSR, larger radius resonators
to achieve efficient overcoupling is to use a pulley-style coupler
\citep{Hosseini2010}. Fig. \ref{fig2}(d) shows the measurement results
for a 100 GHz FSR microresonators coupled with a multi-mode bus waveguide
of the same cross section but in the pulley-style configuration.
The comparison between the two 100 GHz FSR resonators in Fig. \ref{fig2}(b)
and Fig. \ref{fig2}(d) reveals an improved coupling ideality for
the pulley-style coupler, which is however not as high as the case of the 1 THz FSR resonator coupled to a single-mode bus waveguide in Fig \ref{fig2}(c). However such a
comparison neglects the large difference in resonator mode volume.
In fact the fundamental $\mathrm{TM_{R,00}}$ mode of the present
100 GHz FSR resonator can not be overcoupled using a single-mode bus
waveguide, as the strong propagation constant mismatch limits the
achievable coupling rates $\kappa_{\mathrm{ex,0}}$.

\section{Simulations of coupling ideality}

\begin{figure}
\includegraphics[]{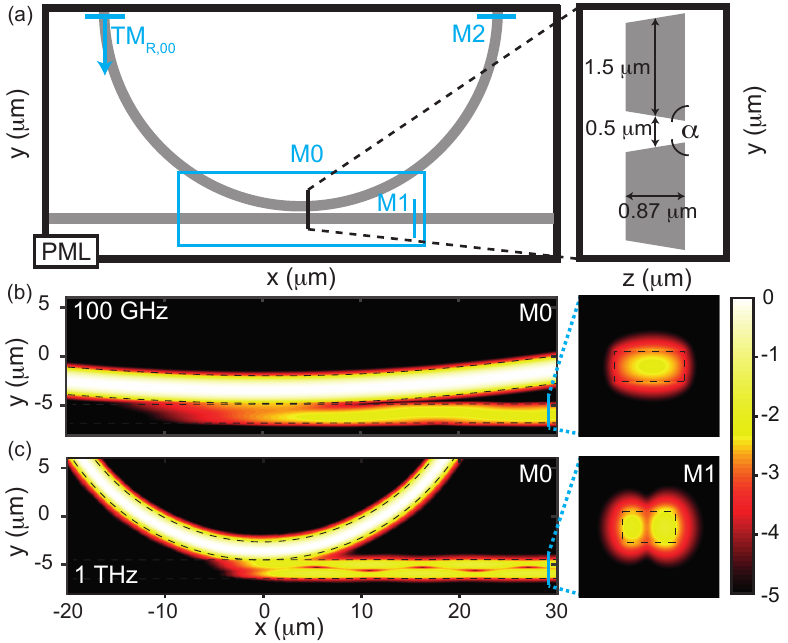}
\caption{FDTD simulations of waveguide coupling for 100 GHz and 1 THz FSR resonators.
(a) Schematic representation of the simulation model. The resonator
and the bus waveguide (both in gray) have the same cross sections
($2.0\times0.87$ $\mathrm{\mu m^2}$ for 100 GHz FSR and $1.5\times0.87$
$\mathrm{\mu m^2}$ for 1 THz FSR) and are separated by $0.5$ $\mathrm{\mu m}$ gap. The sidewall angle is $\alpha=90^{\circ}$. The boundary condition (thick
black lines) enclosing the simulation region is set as PML. The resonator
fundamental $\mathrm{TM_{R,00}}$ mode is launched into the resonator
waveguide and the monitors M0, M1 and M2 record the field distributions
in their individual planes. (b), (c) The field distributions recorded
by M0 and M1 for the 100 GHz and 1 THz FSR resonators. The $\mathrm{TM_{R,00}}$
mode is coupled not only to the bus waveguide fundamental $\mathrm{TM_{B,00}}$
mode but also to its higher-order $\mathrm{TM_{B,10}}$ mode. The
propagation constant difference of both the bus waveguide modes causes
the interference pattern visible along their propagation direction.
This indicates degraded coupling ideality, which is more prominent
in the case of 1 THz FSR. The color bar denotes the field intensity
in logarithmic scale.}
\label{fig3}
\end{figure}

In order to verify the dominant origin of parasitic losses and the
observed strong design-dependence of coupling ideality, we implement
a full 3D finite-difference-time-domain (FDTD) \citep{Taflove2005}
simulation (\textit{Lumerical FDTD}). This allows to study numerically
the light propagation through the coupler by solving Maxwell's equations
in the time domain. The simulation model is shown in Fig. \ref{fig3}(a).
Considering the designs of the microresonator devices experimentally
characterized in the previous section, the resonator and the bus waveguide
have the same cross sections, which is $1.5\times0.87$ $\mathrm{\mu m^{2}}$
(width$\times$height) for the 1 THz FSR resonator and $2.0\times0.87$
$\mathrm{\mu m^{2}}$ for the 100 GHz FSR resonator. The sidewall
angle is $\alpha=90^{\circ}$ and the resonator-bus distance is
set as $0.5$ $\mathrm{\mu m}$. A graded mesh of rectangular cells
with the maximum cell volume of $\mathrm{(22}$ nm$)^{3}$ is applied
to the simulation region. The boundary condition enclosing the full
simulation region is set as perfectly matched layer (PML) \citep{Berenger},
to absorb the incident light to the boundary and thus to prevent back-reflection. 

The resonator fundamental $\mathrm{TM_{R,00}}$ mode at the center
wavelength of 1550 nm is launched with unity power and the light field
propagates until the field distribution reaches the stationary state
in the full simulation region. Monitors M0, M1 and M2 record the field
distributions in their individual monitor planes. Fig. \ref{fig3}(b)
and (c) show the field distributions recorded by M0 and M1 for the
resonators of 100 GHz and 1 THz FSR, respectively. An interference
pattern in the field distribution along the bus waveguide is observed
in both cases, and is more prominent in the case of 1 THz FSR. The
field distributions recorded by M1 show that: (1) in the case of 100
GHz FSR, the field propagates predominantly in the bus waveguide fundamental
$\mathrm{TM_{B,00}}$ mode, which indicates a limited, non-unity coupling
ideality; (2) while in the case of 1 THz FSR, a significant portion
of power is coupled to the higher-order $\mathrm{TM_{B,10}}$ mode
that beats with the $\mathrm{TM_{B,00}}$ mode along the propagation
in the bus waveguide, which indicates a lower coupling ideality. These
qualitative conclusions from Fig. \ref{fig3} agree well with the
experimental observation that the 1 THz FSR resonator in Fig. \ref{fig2}(a)
shows higher parasitic losses thus a lower coupling ideality compared
to the 100 GHz FSR resonator in Fig. \ref{fig2}(b).

\begin{table*}
\begin{tabular}{c|ccccc|cc}
\hline 
No. & 1 & 2 & 3 & 4 & 5 & 6 & 7\tabularnewline
\hline 
FSR & 1 THz & 1 THz & 1 THz & 1 THz & 1 THz & 100 GHz & 100 GHz\tabularnewline
$\mathrm{\mathit{w}_{res}\times\mathit{h}_{res}}$ ($\mathrm{\mu m^{2}}$) & $1.5\times0.87$ & $1.5\times0.87$ & $1.5\times0.87$ & $1.5\times0.87$ & $1.5\times0.87$ & $1.5\times0.87$ & $2.0\times0.87$\tabularnewline
$\mathrm{\mathit{w}_{bus}\times\mathit{h}_{bus}}$ ($\mathrm{\mu m^{2}}$) & - & $1.5\times0.87$ & $1.5\times0.87$ & $1.5\times0.87$ & $0.40\times0.60$ (SM) & $1.5\times0.87$ & $2.0\times0.87$\tabularnewline
gap ($\mu\mathrm{m}$) & - & $0.5$ & $0.2$ & $0.5$ & $0.5$ & $0.5$ & $0.5$\tabularnewline
$\alpha$ & $90^{\circ}$ & $90^{\circ}$ & $90^{\circ}$ & $80^{\circ}$ & $90^{\circ}$ & $90^{\circ}$ & $90^{\circ}$\tabularnewline
\hline 
$\mathrm{\mathit{P}(TM_{R,10})}$ & $1.28\times10^{-4}$ & $7.87\times10^{-4}$ & $0.0116$ & $7.6\times10^{-4}$ & $3.9\times10^{-4}$ & $1.82\times10^{-6}$ & $1.72\times10^{-5}$\tabularnewline
$\mathrm{\mathit{P}(TE_{R,00})}$ & $<10^{-12}$ & $<10^{-12}$ & $<10^{-12}$ & $1.08\times10^{-6}$ & $<10^{-12}$ & $<10^{-12}$ & $<10^{-12}$\tabularnewline
\hline 
$\mathrm{\mathit{P}(TM_{B,00})}$ & - & $3.36\times10^{-3}$ & 0.0344 & $3.31\times10^{-3}$ & $3.92\times10^{-4}$ & 0.0237 & $5.73\times10^{-3}$\tabularnewline
$\mathrm{\mathit{P}(TM_{B,10})}$ & - & 0.0176 & 0.0974 & 0.0173 & - & $3.4\times10^{-6}$ & $1.81\times10^{-4}$\tabularnewline
$\mathrm{\mathit{P}(TE_{B,00})}$ & - & $<10^{-12}$ & $<10^{-12}$ & $4.89\times10^{-6}$ & $<10^{-12}$ & $<10^{-12}$ & $<10^{-12}$\tabularnewline
$\mathrm{\mathit{P}(total)}$ & - & 0.0209 & 0.133 & 0.0203 & $3.92\times10^{-4}$ & 0.0237 & $5.92\times10^{-3}$\tabularnewline
$I$ & - & 0.161 & 0.259 & 0.163 & 1.00 & 1.00 & 0.968\tabularnewline
\hline 
\end{tabular}
\caption{Table of simulated coupled powers for different coupler designs. The
resonator fundamental $\mathrm{TM_{R,00}}$ mode is launched with
unity power. The individual modal powers in the resonator and the
bus waveguides after the coupling section are listed. For every simulated
case the resonator FSR, the cross section of the resonator and the
bus waveguides, the gap distance and the sidewall angle are listed.
$\mathrm{\mathit{P}(total)}$ is the total power recorded in the bus
waveguide after the coupling section and $I$ is the coupling ideality
calculated as $\mathrm{\mathit{I}=\mathit{P}(\text{TM}_{B,00})/\mathit{P}(\text{total})}$.
The bus waveguide is single-mode (SM) in the case No. 5, while all
the other bus waveguides are multi-mode. Modes which do not exist
are marked with hyphen \textquotedbl{} - \textquotedbl{}.}
\label{table1}
\end{table*}

We perform further analysis to quantify the degradation of coupling
ideality in the 1 THz FSR resonators. The total power $\mathrm{\mathit{P}(total)}$
coupled into the bus waveguide can be obtained by calculating the
Poynting vector normal to the monitor plane of M1. In addition, using
the \textquotedbl{}Mode Expansion Function\textquotedbl{} (MEF) of
\textit{Lumerical FDTD}, the field distribution recorded by M1 can
be projected on each waveguide eigenmode and their individual power
($>10^{-12}$) can be calculated. All powers are normalized as they
derive from the resonator fundamental $\mathrm{TM_{R,00}}$ mode that
is launched with unity power. The respective coupling rate $\kappa_{\mathrm{ex},i}$
follows by relating the coupled power to the resonator FSR ($D_{1}/2\pi$)
by $\kappa_{\mathrm{ex},i}=D_{1}\times P(i)$. The fundamental bus
waveguide mode's power $\mathrm{\mathit{P}(TM_{B,00})}$ can be obtained
and the coupling ideality can thus be approximately estimated as $\mathrm{\mathit{I}=\mathit{P}(\text{TM}_{B,00})/\mathit{P}(\text{total})}$,
assuming that the coupling to the higher-order bus waveguide modes
($\kappa_{\mathrm{ex,HOM}}$) is the dominant origin of parasitic
losses. In addition, in order to investigate how the resonator mode
is affected by the coupler, also the field distribution recorded by
M2 in the resonator waveguide after the coupling section is decomposed
into individual resonator modes using MEF.

Table \ref{table1} compiles the simulation results of different coupler
designs (No. 1-7) with varying geometrical parameters, including the
resonator FSR (radius), the cross sections of the resonator and the bus waveguides,
the gap distance, and the waveguide sidewall angle $\alpha$. This
angle $\alpha$ takes into account the fact that the fabricated waveguides have
slanted sidewalls ($\alpha\approx80^{\circ}$). For each design we
calculate the individual power of the selected eigenmodes in the resonator
($\mathrm{TM_{R,10}}$ and transverse electric fundamental resonator
mode $\mathrm{TE_{R,00}}$) and the bus waveguide ($\mathrm{TM_{B,00}}$
,$\mathrm{TM_{B,10}}$ and $\mathrm{TE_{B,00}}$), and numerically
compute the coupling ideality $I$.

First, Table \ref{table1} shows that the commonly employed coupler
design of a straight bus waveguide coupling to a resonator waveguide
of the same cross section, has a higher coupling ideality for the
100 GHz FSR resonators (No. 7, $I\approx0.968$) than for the 1 THz
FSR resonators (No. 2, $I\approx0.161$). This agrees well with the
previously discussed observations in Fig. \ref{fig2} and Fig. \ref{fig3}.
The degraded ideality in the case of 1 THz FSR resonators illustrates
the limited applicability of this coupler design. The fact that the
resonator radius strongly affects coupling ideality is more directly
seen by comparing the cases No. 2 and 6, as both cases have exactly
the same geometrical parameters except for the resonator FSR. 

In addition, the coupling ideality of 100 GHz FSR resonators (No.
6, 7) depends also on the waveguide width when coupled to a
bus waveguide of the same cross section. The degradation of coupling
ideality in the case No. 7 is due to more power coupled to the higher-order
bus waveguide mode ($\mathrm{TM_{B,10}}$), which can be explained
with the smaller propagation constant mismatch between the fundamental
resonator mode ($\mathrm{TM_{R,00}}$) and the higher order bus waveguide
mode ($\mathrm{TM_{B,10}}$). Additionally the wider waveguide cross
section reduces the mutual modal overlap between the fundamental $\mathrm{TM_{R,00}}$
and $\mathrm{TM_{B,00}}$ modes and thus the power transfer
$\mathrm{\mathit{P}(TM_{B,00})}$. Furthermore, our simulations verify
the experimentally observed improvement of coupling ideality for the
1 THz FSR resonator coupled to a single-mode bus waveguide (No. 5,
$I\approx1.00$) . However this is achieved at the expense of reducing
power transfer to the bus waveguide $P(\mathrm{TM_{B,00}})$ by nearly
one order of magnitude, which is due to the propagation constant mismatch
between the $\mathrm{TM_{B,00}}$ and $\mathrm{TM_{R,00}}$ modes.

Second, though only the fundamental $\mathrm{TM_{R,00}}$ mode is
launched in the resonator, a non-zero power in a higher-order mode
$P(\mathrm{TM_{R,10}})$ is recorded by M2. In addition, it is observed
by comparing the uncoupled (No. 1) and coupled cases (No. 2, 3) that
this power in the higher-order resonator mode power $P(\mathrm{TM_{R,10}})$
increases with decreasing gap distance. In the case of the uncoupled
resonator (No. 1), the appearance of $P(\mathrm{TM_{R,10})}=1.28\times10^{-4}$
is mainly attributed to the mesh which acts as a $\mathrm{(22}$ nm$)^{3}$
surface roughness at the material interface. Such surface roughness
is well known to lead to modal coupling e.g. the coupling between
the resonator modes $\mathrm{TM_{R,00}}$ and $\mathrm{TM_{R,10}}$.
In addition, compared with the 100 GHz FSR resonator (No. 6, 7), this
effect is more prominent in the 1 THz FSR resonator (No. 1). Nevertheless
for the coupled resonators (No. 2, 3), the enhancement of $\mathrm{\mathit{P}(TM_{R,10})}$
with decreasing gap distance unambiguously reveals the existence of
a coupler-induced modal coupling. This is an important finding revealing
a novel origin of modal coupling \citep{Herr2014} in microresonators,
which causes distortion of microresonator dispersion properties. 

Third, the coupling of the launched $\mathrm{TM_{R,00}}$ mode to
the modes with the orthogonal polarization, i.e. $\mathrm{TE_{R,00}}$
in the resonator and $\mathrm{TE_{B,00}}$ in the bus waveguide, is
observed in the case of slanted waveguide sidewalls (No. 4). Such
a cross-polarization coupling occurs if the modal field distribution
is asymmetric with respect to its center \citep{Lui1998,Liu2011}
and its strength depends on the degree of this asymmetry. In the simulated
case, the asymmetry is introduced by the ring bending and the $\alpha=80^{\circ}$
sidewall angle. However by comparing the cases No. 2 and 4, the sidewall
angle $\alpha=80^{\circ}$ only enhances significantly the power $P(\mathrm{TE_{R,00}})$,
while the powers of other modes as well as the coupling ideality remain
almost the same.

\section{Conclusion}

In summary we presented the first study of coupling ideality of monolithically
integrated high-Q $\mathrm{Si_{3}N_{4}}$ microresonator devices.
For the commonly employed coupler design where both the resonator
and the bus waveguides have the same cross sections, we revealed the
presence of parasitic losses due to the coupling to higher-order bus
waveguide modes. This degrades coupling ideality which is shown both
through systematic experimental characterization of resonances and
full 3D FDTD simulations. Consequently, an optimized coupler design
using a single-mode bus waveguide with efficiently mitigated parasitic
losses (ideality $I\approx1$) and achieved strong overcoupling ($K>9$)
was demonstrated. Moreover we discovered that the coupler can induce modal coupling between different resonator modes which
is frequently observed in high-Q microresonators.

For microresonator devices based on multi-mode waveguides, coupling
ideality is non-trivial to analyze and strongly depends on coupler
designs and target mode families. In applications, microresonator
devices typically operate around the critical coupling
point, thus high device performance requires optimized coupler designs
with low parasitic losses and high coupling ideality. Our study not
only reveals the design-dependent coupling ideality for integrated
microresonator devices but also demonstrate the importance of anticipating
coupling ideality in device design and the significant improvements
it can unlock.

\subsection*{Acknowledgements}

SiN microresonator samples were fabricated in the EPFL Center of MicroNanotechnology
(CMi). This publication was supported by Contract HR0011-15-C-0055
from the Defense Advanced Research Projects Agency (DARPA), Defense
Sciences Office (DSO) and the Swiss National Science Foundation. M.G.
acknowledges support from the Hasler foundation and support from the
\textquoteleft EPFL Fellows\textquoteright{} fellowship program co-funded
by Marie Curie, FP7 Grant agreement No. 291771.

\bibliographystyle{unsrt}
\bibliography{bibliography}

\begin{thebibliography}{10}

\bibitem{Little1997}
B.~E. Little, S.~T. Chu, H.~A. Haus, J.~Foresi, and J.~Laine.
\newblock Microring resonator channel dropping filters.
\newblock {\em J. Lightwave Technol.}, 15(6):998--1005, 1997.

\bibitem{DeVos2007}
K.~{De Vos}, I.~Bartolozzi, E.~Schacht, P.~Bienstman, and R.~Baets.
\newblock Silicon-on-insulator microring resonator for sensitive and label-free
  biosensing.
\newblock {\em Opt. Express}, 15(12):7610--7615, 2007.

\bibitem{Reed2005}
G.~T. Reed, G.~Mashanovich, F.~Y. Gardes, and D.~J. Thomson.
\newblock Silicon optical modulators.
\newblock {\em Nat. Photon.}, 4(8):518--526, 08 2010.

\bibitem{Leuthold2010}
J.~Leuthold, C.~Koos, and W.~Freude.
\newblock Nonlinear silicon photonics.
\newblock {\em Nat. Photon.}, 4(8):535--544, 08 2010.

\bibitem{Moss2013}
D.~J. Moss, R.~Morandotti, A.~L. Gaeta, and M.~Lipson.
\newblock New {CMOS}-compatible platforms based on silicon nitride and hydex
  for nonlinear optics.
\newblock {\em Nat. Photon.}, 7(8):597--607, 2013.

\bibitem{Li2015}
Q.~Li, M.~Davan{\c c}o, and K.~Srinivasan.
\newblock Efficient and low-noise single-photon-level frequency conversion
  interfaces using silicon nanophotonics.
\newblock {\em Nat. Photon.}, 10(6):406--414, 06 2016.

\bibitem{Kippenberg2011}
T.~J. Kippenberg, R.~Holzwarth, and S.~A. Diddams.
\newblock Microresonator-based optical frequency combs.
\newblock {\em Science}, 332(6029):555--559, 2011.

\bibitem{Jalali2006}
B.~Jalali and S.~Fathpour.
\newblock Silicon photonics.
\newblock {\em J. Lightwave Technol.}, 24(12):4600--4615, 2006.

\bibitem{Nagarajan2005}
R.~Nagarajan, C.~H. Joyner, R.~P. Schneider, J.~S. Bostak, T.~Butrie, A.~G.
  Dentai, V.~G. Dominic, P.~W. Evans, M.~Kato, M.~Kauffman, D.~J.~H. Lambert,
  S.~K. Mathis, A.~Mathur, R.~H. Miles, M.~L. Mitchell, M.~J. Missey,
  S.~Murthy, A.~C. Nilsson, F.~H. Peters, S.~C. Pennypacker, J.~L. Pleumeekers,
  R.~A. Salvatore, R.~K. Schlenker, R.~B. Taylor, H.-S. Tsai, M.~F. {Van
  Leeuwen}, J.~Webjorn, M.~Ziari, D.~Perkins, J.~Singh, S.~G. Grubb, M.~S.
  Reffle, D.~G. Mehuys, F.~A. Kish, and D.~F. Welch.
\newblock Large-scale photonic integrated circuits.
\newblock {\em IEEE J. Sel. Top. Quantum Electron.}, 11(1):50--65, 2005.

\bibitem{Cherchi2013}
M.~Cherchi, S.~Ylinen, M.~Harjanne, M.~Kapulainen, and T.~Aalto.
\newblock Dramatic size reduction of waveguide bends on a micron-scale silicon
  photonic platform.
\newblock {\em Opt. Express}, 21(15):17814--17823, 2013.

\bibitem{Bauters2011}
J.~F. Bauters, M.~J.~R. Heck, D.~D. John, J.~S. Barton, C.~M. Bruinink,
  A.~Leinse, R.~G. Heideman, D.~J. Blumenthal, and J.~E. Bowers.
\newblock Planar waveguides with less than 0.1 d{B}/m propagation loss
  fabricated with wafer bonding.
\newblock {\em Opt. Express}, 19(24):24090--24101, 2011.

\bibitem{Luo2014}
L.-W. Luo, N.~Ophir, C.~P. Chen, L.~H. Gabrielli, C.~B. Poitras, K.~Bergmen,
  and M.~Lipson.
\newblock {WDM}-compatible mode-division multiplexing on a silicon chip.
\newblock {\em Nat. Commun.}, 5(3069), 2014.

\bibitem{Wade2014}
M.~T. Wade, J.~M. Shainline, J.~S. Orcutt, R.~J. Ram, V.~Stojanovic, and M.~A.
  Popovic.
\newblock Spoked-ring microcavities: enabling seamless integration of
  nanophotonics in unmodified advanced {CMOS} microelectronics chips.
\newblock {\em Proc. SPIE}, 8991(303):89910B--89910B--8, 2014.

\bibitem{Turner2006}
A.~C. Turner, C.~Manolatou, B.~S. Schmidt, M.~Lipson, M.~A. Foster, J.~E.
  Sharping, and A.~L. Gaeta.
\newblock Tailored anomalous group-velocity dispersion in silicon channel
  waveguides.
\newblock {\em Opt. Express}, 14(10):4357--4362, 2006.

\bibitem{Riemensberger2012}
J.~Riemensberger, K.~Hartinger, T.~Herr, V.~Brasch, R.~Holzwarth, and T.~J.
  Kippenberg.
\newblock Dispersion engineering of thick high-{Q} silicon nitride
  ring-resonators via atomic layer deposition.
\newblock {\em Opt. Express}, 20(25):27661--27669, 2012.

\bibitem{Braginsky1989}
V.~B. Braginsky, M.~L. Gorodetsky, and V.~S. Ilchenko.
\newblock Quality-factor and nonlinear properties of optical whispering-gallery
  modes.
\newblock {\em Phys. Lett. A}, 137(7):393--397, 1989.

\bibitem{Knight1997}
J.~C. Knight, G.~Cheung, F.~Jacques, and T.~A. Birks.
\newblock Phase-matched excitation of whispering-gallery-mode resonances by a
  fiber taper.
\newblock {\em Opt. Letters}, 22(15):1129--1131, 1997.

\bibitem{Gorodetsky1999}
M.~L. Gorodetsky and V.~S. Ilchenko.
\newblock Optical microsphere resonators: optimal coupling to high-{Q}
  whispering gallery modes.
\newblock {\em J. Opt. Soc. Am. B}, 16(1):147--164, 1999.

\bibitem{Cai2000}
M.~Cai, O.~Painter, and K.~J. Vahala.
\newblock Observation of critical coupling in a fiber taper to a
  silica-microsphere whispering-gallery mode system.
\newblock {\em Phys. Rev. Lett.}, 85:74--77, 2000.

\bibitem{Spillane2003}
S.~M. Spillane, T.~J. Kippenberg, O.~J. Painter, and K.~J. Vahala.
\newblock Ideality in a fiber-taper-coupled microresonator system for
  application to cavity quantum electrodynamics.
\newblock {\em Phys. Rev. Lett.}, 91:043902, 2003.

\bibitem{Soltani2007}
M.~Soltani, S.~Yegnanarayanan, and A.~Adibi.
\newblock Ultra-high {Q} planar silicon microdisk resonators for chip-scale
  silicon photonics.
\newblock {\em Opt. Express}, 15(8):4694--4704, 2007.

\bibitem{Hosseini2010}
E.~S. Hosseini, S.~Yegnanarayanan, A.~H. Atabaki, M.~Soltani, and A.~Adibi.
\newblock Systematic design and fabrication of high-{Q} single-mode
  pulley-coupled planar silicon nitride microdisk resonators at visible
  wavelengths.
\newblock {\em Opt. Express}, 18(3):2127--2136, 2010.

\bibitem{Ghulinyan2013}
M.~Ghulinyan, F.~Ramiro-Manzano, N.~Prtljaga, R.~Guider, I.~Carusotto,
  A.~Pitanti, G.~Pucker, and L.~Pavesi.
\newblock Oscillatory vertical coupling between a whispering-gallery resonator
  and a bus waveguide.
\newblock {\em Phys. Rev. Lett.}, 110(16):163901, 2013.

\bibitem{Pencer2014}
D.~T. Spencer, J.~F. Bauters, M.~J.~R. Heck, and J.~E. Bowers.
\newblock Integrated waveguide coupled {S}i3{N}4 resonators in the
  ultrahigh-{Q} regime.
\newblock {\em Optica}, 1(3):153--157, 2014.

\bibitem{Fengnian2006}
F.~Xia, L.~Sekaric, and Y.~A. Vlasov.
\newblock Mode conversion losses in silicon-on-insulator photonic wire based
  racetrack resonators.
\newblock {\em Opt. Express}, 14(9):3872--3886, 2006.

\bibitem{Li2016}
A.~Li, T.~{Van Vaerenbergh}, P.~{De Heyn}, P.~Bienstman, and W.~Bogaerts.
\newblock Backscattering in silicon microring resonators: a quantitative
  analysis.
\newblock {\em Laser Photon. Rev.}, 431:0--13, 2016.

\bibitem{Herr2014}
T.~Herr, V.~Brasch, J.~D. Jost, I.~Mirgorodskiy, G.~Lihachev, M.~L. Gorodetsky,
  and T.~J. Kippenberg.
\newblock Mode spectrum and temporal soliton formation in optical
  microresonators.
\newblock {\em Phys. Rev. Lett.}, 113:123901, 2014.

\bibitem{Braschaad4811}
V.~Brasch, M.~Geiselmann, T.~Herr, G.~Lihachev, M.~H.~P. Pfeiffer, M.~L.
  Gorodetsky, and T.~J. Kippenberg.
\newblock Photonic chip{\textendash}based optical frequency comb using soliton
  {C}herenkov radiation.
\newblock {\em Science}, 351(6271):357--360, 2016.

\bibitem{haus1984waves}
H.~A. Haus.
\newblock {\em Waves and Fields in Optoelectronics}.
\newblock Prentice-Hall Series in Solid State Physical Electronics. Prentice
  Hall, Incorporated, 1984.

\bibitem{Rowland1993}
D.~R. Rowland and J.~D. Love.
\newblock Evanescent wave coupling of whispering gallery modes of a dielectric
  cylinder.
\newblock {\em IEE Proceedings J-Optoelectronics}, 140(3):177, 1993.

\bibitem{Yariv2000}
A.~Yariv.
\newblock Universal relations for coupling of optical power between
  microresonators and dielectric waveguides.
\newblock {\em Electron. Lett.}, 36(4):321--322, 2000.

\bibitem{Yariv1973}
A~Yariv.
\newblock Coupled-mode theory for guided-wave optics.
\newblock {\em IEEE J. Quant. Electron.}, 9(9):919--933, 1973.

\bibitem{Almeida2003}
V.~R. Almeida, R.~R. Panepucci, and M.~Lipson.
\newblock Nanotaper for compact mode conversion.
\newblock {\em Opt. Lett.}, 28(15):1302--1304, 2003.

\bibitem{Herr2013a}
T.~Herr, V.~Brasch, J.~D. Jost, C.~Y. Wang, N.~M. Kondratiev, M.~L. Gorodetsky,
  and T.~J. Kippenberg.
\newblock Temporal solitons in optical microresonators.
\newblock {\em Nat. Photon.}, 8(1):145--152, 2014.

\bibitem{Pfeiffer2016}
M.~H.~P. Pfeiffer, A.~Kordts, V.~Brasch, M.~Zervas, M.~Geiselmann, J.~D. Jost,
  and T.~J. Kippenberg.
\newblock Photonic {D}amascene process for integrated high-{Q} microresonator
  based nonlinear photonics.
\newblock {\em Optica}, 3(1):20--25, 2016.

\bibitem{DelHaye2009}
P.~Del'Haye, O.~Arcizet, M.~L. Gorodetsky, R.~Holzwarth, and T.~J. Kippenberg.
\newblock Frequency comb assisted diode laser spectroscopy for measurement of
  microcavity dispersion.
\newblock {\em Nat. Photon.}, 3(9):529--533, 2009.

\bibitem{Gorodetsky2000}
M.~L. Gorodetsky, A.~D. Pryamikov, and V.~S. Ilchenko.
\newblock Rayleigh scattering in high-{Q} microspheres.
\newblock {\em J. Opt. Soc. Am. B}, 17(6):1051--1057, 2000.

\bibitem{Taflove2005}
A.~Taflove and S.~C. Hagness.
\newblock {\em Computational Electrodynamics: The Finite-Difference Time-Domain
  Method}.
\newblock Artech House antennas and propagation library. Artech House, 2005.

\bibitem{Berenger}
J.-P. Berenger.
\newblock {\em Perfectly Matched Layer (PML) for Computational
  Electromagnetics}.
\newblock Morgan $\&$ Claypool, 2007.

\bibitem{Lui1998}
W.~W. Lui, T.~Hirono, K.~Yokoyama, and W.-P. Huang.
\newblock Polarization rotation in semiconductor bending waveguides: a
  coupled-mode theory formulation.
\newblock {\em J. Lightwave Technol.}, 16(5):929--936, 1998.

\bibitem{Liu2011}
L.~Liu, Y.~Ding, K~Yvind, and J.~M. Hvam.
\newblock Efficient and compact {TE}--{TM} polarization converter built on
  silicon-on-insulator platform with a simple fabrication process.
\newblock {\em Opt. Lett.}, 36(7):1059--1061, 2011.

\end{thebibliography}

\end{document}